\documentclass[preprint,superscriptaddress,amsmath,amssymb,aps,pre,a4paper,floatfix]{revtex4-2}

\usepackage[T1]{fontenc}
\usepackage[utf8]{inputenc}
\usepackage{bm}
\usepackage{booktabs}
\usepackage{siunitx}
\usepackage{graphicx}
\usepackage{placeins}
\usepackage{hyperref}
\hypersetup{colorlinks=true,linkcolor=blue,citecolor=blue,urlcolor=blue}
\begin{document}

\title{Polar Phase of an Egg-Yolk Particle in a Rotating Magnetic Field}

\author{Roberts Treize}
\author{J\={a}nis C\={\i}murs}
\email{janis.cimurs@lu.lv}
\affiliation{Laboratory of Magnetic Soft Materials, Physics Department,
Faculty of Science and Technology, University of Latvia,
Jelgavas iela 3, Riga, LV-1002, Latvia}

\date{\today}

\begin{abstract}
The orbit selected by a magnetic particle in a viscous liquid under a
rotating field is commonly described in the instantaneous
Stoner--Wohlfarth limit, in which the magnetic moment always occupies a
minimum of the anisotropy energy. In this limit, precession out of the
field plane exists only for reduced fields $B_0/B_a<1/\sqrt2$, and
stronger fields admit only planar synchronous rotation or libration. We
study a macroscopic egg-yolk particle, a printed shell containing two
freely rotating permanent magnets in fluid-filled cavities, driven by a
compensated rotating field and tracked optically. In addition to
synchronous rotation, precession, and libration, the particle exhibits a
polar phase: above the instantaneous bound and at high frequency, the
shell axis leaves the field plane and locks onto a narrow cone about the
rotation axis. We extend the model by a finite viscous friction between
the magnets and the shell. The extension introduces a single
dimensionless parameter, the ratio $\kappa$ of inner to outer rotational
friction, and reduces to the instantaneous theory as $\kappa\to0$.
Numerical phase diagrams show that increasing $\kappa$ tilts the bistable
band to higher fields at higher frequencies and extends stable precession
above $B_0/B_a=1/\sqrt2$, whereas for large $\kappa$ planar libration
returns at high frequency. With a calibrated anisotropy field, relaxation
rate, and $\kappa$ of order unity, the model reproduces the terminal
state of all measured records and their tilt angles within a few
degrees, including the polar cone. In the model the polar state is
maintained by a steady rotation of the magnets relative to the shell,
driven by the unbalanced field and anisotropy torques and dissipated by
the inner friction; it therefore exists only for intermediate $\kappa$.
Internal friction thus acts as a control parameter for orbit selection
in magnetic particles with an internal rotational degree of freedom.
\end{abstract}


\maketitle
\raggedbottom

\section{Introduction}

A magnetic particle in a viscous liquid is a torque balance among the
applied field, the magnetic anisotropy, and the hydrodynamic drag
\cite{rosensweig1985,usov2012,shliomis1993,helbig2023}. When the field
rotates, that balance produces a small number of robust orbits: the
body may lock to the field, slip in the field plane, or leave the plane
and precess \cite{brics2023,stikuts2023,boniface2024}. Which orbit is
selected is what one actually observes and controls in practice, from
the rotation of magnetotactic bacteria
\cite{erglis2007,satyanarayana2021,mirkhani2024,smite2025,villanueva2025}
to the torque of lab-on-chip rotors
\cite{Moerland2019,ranzoni2010magnetically,raghu2026} and the
synchronization of colloids \cite{yan2012,domingos2025,tavacoli2025}.
At the nanoscale, the magnetic direction can also reverse on its own,
by thermal N\'eel--Brown activation
\cite{neel1949,brown1963,ilg2024,davidson2024} or stochastic
Landau--Lifshitz--Gilbert dynamics
\cite{garciaPalacios1998,taukulis2012,mostarac2025}. For a macroscopic
moment, those channels are frozen, and what remains is conservative
Stoner--Wohlfarth remagnetization \cite{stonerWohlfarth} coupled to
the overdamped rotation of the body. How much of the observed orbit
selection this minimal description explains is the question of this
article.

C\={\i}murs and C\={e}bers \cite{cimurs2013PRE,cimurs2013PRE87}
answered it in the limit where the magnetic moment responds
instantaneously: the internal magnetic direction $\mathbf e$ always
sits in a local minimum of the anisotropy energy and jumps when that
minimum disappears, so that only the body axis $\mathbf n$ has its own
dynamics. Two numbers then control the motion: the field amplitude
relative to the anisotropy field, $\hat B_0=B_0/B_a$, and the field
frequency relative to the viscous relaxation rate,
$\hat{\omega}=\omega/\omega_a$. The theory predicts four kinds of
motion: rotation in step with the field; precession on a cone, possible
only at weak fields, $\hat B_0<1/\sqrt2$; back-and-forth libration in
the field plane, with repeated slips of about $180^\circ$; and a
bistable region where the initial condition decides between the planar
and the conical state. Its prediction for strong, fast fields is
definite: the body stays in the field plane, and no state exists in
which $\mathbf n$ rises toward the rotation axis.

That instantaneous picture is an idealization. In a real particle the
magnetization, or the embedded magnets, rub against the body, so
$\mathbf e$ follows $\mathbf n$ and $\mathbf B$ only after a finite
relaxation time and can lag behind both. This internal friction is
hard to see in a colloid, where usually only one optical axis is
available. Macroscopic analogues make it observable
\cite{SPULIS2024171647,teixeira2024}. The egg-yolk particle used here
is a $14$-mm printed shell holding two freely rotating N45 magnets in
glycerol-filled cavities \cite{treize2026}. The magnets turn inside the
shell with their own finite response time, so the particle realizes
exactly the finite-time remagnetization that the instantaneous theory
leaves out. The black and white hemispheres make the shell axis
$\mathbf n$ visible to a single camera, the field is compensated in
three axes, and thermal fluctuations are negligible.

The same particle has already shown synchronous rotation, precession,
and libration \cite{treize2026}. The new observation, and the subject
of this article, is a polar phase that appears once the field is
driven at higher frequency: instead of librating in the field plane,
the shell axis lifts out of the plane and settles into a steady
precession around the rotation axis, standing nearly upright. The
instantaneous theory has no such state; it predicts planar libration
throughout that regime. To find out whether the finite response of the
magnets is responsible, we extend the model by a single inner friction
$\gamma_{\mathrm{in}}$ between the magnets and the shell. This keeps
$\mathbf e$ as a dynamical variable, adds one dimensionless parameter,
the friction ratio $\kappa=\gamma_{\mathrm{in}}/\gamma_{\mathrm{out}}$,
and recovers the instantaneous theory as $\kappa\to0$. We then compare
the measured motion with both the instantaneous and the two-friction
model at the laboratory reference frame.

The article is organized as follows. Section~\ref{sec:experiment}
describes the particle, the field platform, the observables, and the
measured runs. Section~\ref{sec:model} formulates the two-friction
model and its limits, and Sec.~\ref{sec:numerics} gives the numerical
protocol. Section~\ref{sec:results} presents the phase diagram as a
function of $\kappa$, the calibration of the model against the
experiment, the time series of the four measured regimes, and the
structure of the polar state. Section~\ref{sec:discussion} discusses
the agreement, the calibrated scales, and the open questions, and concludes.
\section{Experimental foundation}
\label{sec:experiment}

\subsection{Particle and field platform}

The particle consists of two black and white PLA hemispheres with total
diameter $D=\SI{14}{\milli\metre}$ and $95\%$ infill. A cross-shaped lock
prevents relative rotation of the hemispheres. Each half contains one N45
NdFeB sphere of diameter \SI{3.0}{\milli\metre} with moment
$m_0=\SI{0.012}{\ampere\metre\squared}$. The magnets sit in
glycerol-filled cavities with a clearance of about
\SI{0.2}{\milli\metre} and are free to rotate; their center separation
is $d=\SI{7}{\milli\metre}$. The total moment is
$m=2m_0=\SI{0.024}{\ampere\metre\squared}$.\cite{treize2026}

For two parallel dipoles on the shell axis, the orientation-dependent
interaction has the uniaxial form
\begin{equation}
E_{dd}=-\frac{\mu_0m_0^2}{4\pi d^3}
       \left[3(\mathbf e\cdot\mathbf n)^2-1\right]
       \equiv-\frac{KV}{2}(\mathbf e\cdot\mathbf n)^2+\mathrm{const.}
\label{eq:dipole}
\end{equation}
The dipolar form is exact for uniformly magnetized spheres, and the
geometry gives $KV/2=3\mu_0m_0^2/(4\pi d^3)=\SI{0.13}{\milli\joule}$.
The shell rotates in a rectangular pocket filled with glycerol of nominal viscosity
$\eta=\SI{1000}{\milli\pascal\second}$. For a sphere in unbounded fluid
the rotational drag is $8\pi\eta(D/2)^3=\SI{0.86e-5}{\joule\second}$; a
confinement correction based on low-Reynolds-number hydrodynamics raises
it to $\gamma_{\mathrm{out}}\simeq\SI{1e-5}{\joule\second}$
\cite{happel1983,rosensweig1985}. The model below is written in terms of
the Stoner--Wohlfarth anisotropy field and the viscous relaxation rate
of the pair,
\begin{align}
B_a&=\frac{KV}{m}=\SI{10.5}{\milli\tesla},
\label{eq:Ba}\\
\omega_a&=\frac{KV}{\gamma_{\mathrm{out}}}\approx\SI{25}{\per\second}
       \;(\SI{4.0}{\hertz}).
\label{eq:wa}
\end{align}
In these units the smallest switching field of the pair is $B_a/2$
\cite{stonerWohlfarth}, and the largest rate at which the anisotropy
torque, at most $KV/2$, can turn the shell is $\omega_a/2$.

\begin{figure}[!htbp]
\centering
\includegraphics[width=0.9\columnwidth]{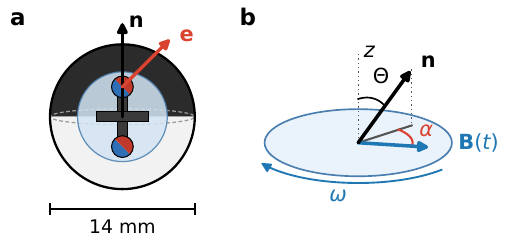}
\caption{\textbf{a}, Egg-yolk particle: 14-mm PLA shell and two free
inner N45 magnets. The shell axis is $\mathbf n$ and the internal
magnetic direction is $\mathbf e$. \textbf{b}, Polar angle
$\Theta=\arccos n_z$ and lag $\alpha$ between the field $\mathbf B(t)$
and the projection of $\mathbf n$ on the field plane. $\Theta=90^\circ$
is the field-plane equator (planar libration in the instantaneous
theory); $\Theta\to 0$ is the $+z$ pole, the new high-drive phase.}
\label{fig:platform}
\end{figure}

Three orthogonal Helmholtz pairs generate the field. Two channels,
driven by KEPCO power supplies, produce the rotating field
\begin{equation}
\mathbf B(t)=B_0(\cos2\pi ft,\,\sin2\pi ft,\,0),
\label{eq:field}
\end{equation}
and the local Earth field, approximately \SI{51.2}{\micro\tesla}, is
compensated to within \SI{0.2}{\micro\tesla}. The
edge series of Sec.~\ref{sec:calibration} is converted with a separately
fixed factor. A Basler acA1920 camera with a
\SI{35}{\milli\metre} lens records the particle for \SI{20}{\second} at
30 frames per second (run d: \SI{10}{\second} at 25 frames per second).
The tracking workflow thresholds the black and white hemispheres and
synchronizes each orientation estimate with the coil currents
\cite{treizeToolkit2025}. The apparatus is that of Ref.~\cite{treize2026}: a three-axis Helmholtz coil
system (inner coil diameters 21.5, 29.4, and 39.0~cm for the $x$, $y$, and $z$
pairs) driven by three bipolar power supplies (Kepco BOP~20-10M, $\pm20$~V,
$\pm10$~A) under real-time LabVIEW control, with the Earth's field compensated
and verified by an HMC5883L magnetometer. The particle rests in the central
pocket of a transparent PLA container filled with glycerol, illuminated from
above by LED strips, and is imaged from above along $z$ by a monochrome camera
(Basler acA1920-155um, 35~mm lens, 30~frames\,s$^{-1}$). Figure~\ref{fig:platform} shows the particle
and the angles used throughout this article.

\subsection{Observables and angle conventions}
\label{sec:observables}

The orientation of the shell axis $\mathbf n$ is described by two
angles, Fig.~\ref{fig:platform}(b), with the conventions of
Ref.~\cite{treize2026}. The field rotates in the $xy$ plane about the
$z$ axis with angular frequency $\omega=2\pi f$. The polar angle
\begin{equation}
\Theta=\arccos n_z
\label{eq:theta}
\end{equation}
measures the tilt of $\mathbf n$ from the rotation axis: $\Theta=90^\circ$
is motion in the field plane and $\Theta=0$ is alignment with the $+z$
axis. Cones about $+z$ and $-z$ are mirror images of each other
(Sec.~\ref{sec:symmetry}), and we report the folded tilt
\begin{equation}
\Theta_{\mathrm{fold}}=\arccos|n_z|=\min(\Theta,\,180^\circ-\Theta);
\label{eq:thetafold}
\end{equation}
from here on, quoted tilts are $\Theta_{\mathrm{fold}}$ and are written
$\Theta$. The camera records the particle from the $+z$ direction. When $\mathbf{n}$ is near the
pole the single-view estimate, which is based on the visible black--white
boundary of the particle, is biased low.

The lag angle
\begin{equation}
\alpha(t)=\omega t-\varphi_n(t),\qquad
\varphi_n=\operatorname{atan2}(n_y,n_x),
\label{eq:alpha}
\end{equation}
is the azimuth of the field minus the azimuth of the projection of
$\mathbf n$ onto the field plane. It is accumulated continuously in
time. Its late-time slope defines the slip
$s=\langle\dot\alpha\rangle/\omega$, the number of field turns lost per
turn: $s=0$ for synchronous rotation and for a cone locked to the
field, and $s>0$ for libration, where $\alpha$ grows in slips of about
$180^\circ$.

Working points are given in the reduced coordinates of the
instantaneous theory \cite{cimurs2013PRE},
\begin{equation}
\hat{\omega}=\frac{\omega}{\omega_a},\qquad
\hat{B}_0=\frac{B_0}{B_a},
\label{eq:xy}
\end{equation}
with $B_a$ and $\omega_a$ from Eqs.~\eqref{eq:Ba} and \eqref{eq:wa}, or
from the calibration of Sec.~\ref{sec:calibration}.

Thermal fluctuations are negligible: the Langevin parameter
$mB_0/k_BT$ exceeds $10^{16}$, in contrast to the stochastic dynamics of
single-domain nanoparticles
\cite{taukulis2012,garciaPalacios1998,pyanzina2025}. Inertia is small
but not negligible at the highest rates. With glycerol density
$\rho\approx\SI{1.26}{\gram\per\centi\metre\cubed}$, the rotational
Reynolds number of the shell is
$\mathrm{Re}=\rho\,\Omega(D/2)^2/\eta\approx(\SI{0.06}{\second})\,\Omega$
for a shell angular velocity $\Omega$: it is below unity in the
synchronous and precessing runs at \SIrange{1.8}{2.2}{\hertz} and of
order unity in the polar state at \SI{8}{\hertz}. The inertial number of
the shell, $I_s\omega/\gamma_{\mathrm{out}}$ with
$I_s\approx\SI{3e-8}{\kilogram\metre\squared}$, is about $0.04$ at
\SI{2.2}{\hertz} and $0.15$ at \SI{8}{\hertz}. The model of
Sec.~\ref{sec:model} is overdamped and neglects both; we return to
their effect in Sec.~\ref{sec:discussion}.

The reduced coordinates carry systematic uncertainties larger than the
instrument readout errors. $B_a\propto m_0/d^3$ is sensitive to the
tolerance of the nominal moment and, through $d^{-3}$, to the positions
of the magnets in their cavities, and Eq.~\eqref{eq:dipole} assumes that
the two moments stay parallel. $\omega_a$ depends in addition on
$\gamma_{\mathrm{out}}$, which is inferred from a confinement correction for an
idealized cavity, whereas the actual pocket is rectangular. An error in
$KV$ therefore shifts a working point along both axes of the
$(\hat{\omega},\hat{B}_0)$ plane, an error in $\gamma_{\mathrm{out}}$ shifts it along
$\hat{\omega}$ only, and an error in the coil constant shifts it along
$\hat B_0$ only. The reduced coordinates of the measured working points
are consequences of these estimates, not independent calibrations of
the analytical phase boundaries.

\subsection{Measured runs}
\label{sec:runs}

We demonstrate the agreement between the model and the experiment with
four runs, labeled a--d (Table~\ref{tab:runs} and
Sec.~\ref{sec:results}).
Runs a--c were reported in Ref.~\cite{treize2026} and are reanalyzed
here with the conventions above; run d is new.
(a)~At \SI{11.2}{\milli\tesla} and \SI{1.8}{\hertz} the shell rotates
synchronously in the field plane with a bounded lag.
(b)~At \SI{8.0}{\milli\tesla} and \SI{2.2}{\hertz} it leaves the plane
and precesses on a cone locked to the field, at $\Theta\simeq67^\circ$.
(c)~At \SI{11.2}{\milli\tesla} and \SI{2.2}{\hertz} it stays in the
plane, $\Theta\simeq86^\circ$, and librates; the lag grows in slips of
about $180^\circ$.
(d)~At \SI{15.8}{\milli\tesla} and \SI{8}{\hertz} it starts near the
plane, rises within about two seconds, and locks onto a cone
\SIrange{20}{24}{\degree} from the rotation axis. This is the polar
phase.
In addition, the edge of the libration region was recorded between
\SI{4.0}{\hertz} and \SI{7.3}{\hertz} (Sec.~\ref{sec:calibration}).
This edge, the black line in Fig.~\ref{fig:phase_kappa}, is the measured
boundary between planar libration and the polar precession, and it was
obtained by stepping the drive rather than by a continuous sweep. The
field amplitude was held fixed and the frequency was raised in steps
of \SI{0.1}{\hertz}, 
each step being held until the motion had settled. The first frequency
at which the shell axis left the field plane and stayed on a cone
around the rotation axis was recorded as the edge of the libration
region at that amplitude. Repeating this at five amplitudes gave the
points (\SI{4.0}{\hertz}, \SI{10.2}{\milli\tesla}),
(\SI{5.6}{\hertz}, \SI{11.4}{\milli\tesla}),
(\SI{6.8}{\hertz}, \SI{13.7}{\milli\tesla}),
(\SI{7.1}{\hertz}, \SI{15.9}{\milli\tesla}), and
(\SI{7.3}{\hertz}, \SI{18.2}{\milli\tesla}), which are joined by the
black line; since the boundary was always approached from the
libration side, these are the frequencies at which the planar state
loses stability to the polar one.
\section{Theory}
\label{sec:model}

The particle is a rigid shell that encloses a pair of permanent magnets
in fluid-filled cavities. Its state is described by two unit vectors:
the shell axis $\mathbf n$ and the internal magnetic direction
$\mathbf e$,
\begin{equation}
|\mathbf n|=1,\qquad |\mathbf e|=1.
\label{eq:unit}
\end{equation}
The shell rotates against the outer glycerol, and the magnets rotate
relative to the shell against the inner fluid. We treat both rotations
as overdamped, each with its own friction coefficient. In the limit of
vanishing inner friction the model reduces to the instantaneous
Stoner--Wohlfarth description of
Refs.~\cite{cimurs2013PRE,cimurs2013PRE87}.

\subsection{Kinematics, energy, and torques}

Both vectors move on the unit sphere by rigid-body kinematics,
\begin{equation}
\dot{\mathbf n}=\bm\Omega_n\times\mathbf n,\qquad
\dot{\mathbf e}=\bm\Omega_e\times\mathbf e,
\label{eq:kinematics}
\end{equation}
with $\bm\Omega_e=\bm\Omega_n+\bm\omega_{\mathrm{rel}}$, where
$\bm\omega_{\mathrm{rel}}$ is the spin of the magnets relative to the
shell.

The magnetic energy of the inner pair in an applied field is
\begin{equation}
E=-m\mathbf e\cdot\mathbf B
  -\frac{KV}{2}(\mathbf e\cdot\mathbf n)^2,
\label{eq:energy}
\end{equation}
with $KV/2$ from Eq.~\eqref{eq:dipole}.
A virtual rotation $\delta\mathbf e=\delta\bm\varphi_e\times\mathbf e$,
$\delta\mathbf n=\delta\bm\varphi_n\times\mathbf n$ gives the
torques
\begin{align}
\bm{\mathcal T}_e
&=m\mathbf e\times\mathbf B
 +KV(\mathbf e\cdot\mathbf n)\,(\mathbf e\times\mathbf n),
\label{eq:Te}\\
\bm{\mathcal T}_n
&=KV(\mathbf e\cdot\mathbf n)\,(\mathbf n\times\mathbf e)
=-\,KV(\mathbf e\cdot\mathbf n)\,(\mathbf e\times\mathbf n).
\label{eq:Tn}
\end{align}
The anisotropy contributions are equal and opposite: they redistribute
angular momentum between the magnets and the shell and do no net work
on the composite. The external field torque $m\mathbf e\times\mathbf B$
is the only couple that can turn the particle against the outer fluid.

\subsection{Two-friction dynamics}

With $\gamma_{\mathrm{out}}$ the rotational friction of the shell
against the outer glycerol (Sec.~\ref{sec:experiment}) and
$\gamma_{\mathrm{in}}$ that of the magnets in their cavities, the
Rayleigh dissipation function is
\begin{equation}
\mathcal R=\frac12\gamma_{\mathrm{out}}|\bm\Omega_n|^2
          +\frac12\gamma_{\mathrm{in}}|\bm\Omega_e-\bm\Omega_n|^2.
\label{eq:rayleigh}
\end{equation}
Imposing $\partial\mathcal R/\partial\bm\Omega_e=\bm{\mathcal T}_e$
and $\partial\mathcal R/\partial\bm\Omega_n=\bm{\mathcal T}_n$ gives
$\gamma_{\mathrm{in}}\bm\omega_{\mathrm{rel}}=\bm{\mathcal T}_e$ and
$\gamma_{\mathrm{out}}\bm\Omega_n=\bm{\mathcal T}_e+\bm{\mathcal T}_n$.
The anisotropy terms cancel in the second identity, and one obtains
\begin{align}
\bm\Omega_n
&=\frac{m}{\gamma_{\mathrm{out}}}\,\mathbf e\times\mathbf B,
\label{eq:omega-n}\\
\bm\omega_{\mathrm{rel}}
&=\frac{1}{\gamma_{\mathrm{in}}}
\bigl[m\mathbf e\times\mathbf B
 +KV(\mathbf e\cdot\mathbf n)\,(\mathbf e\times\mathbf n)\bigr].
\label{eq:omega-rel}
\end{align}
Thus the shell is driven only by the field torque, while both the
field and the anisotropy act on the relative spin
\cite{usov2012,shliomis1993}. The kinematic closure is
\begin{equation}
\dot{\mathbf n}=\bm\Omega_n\times\mathbf n,\qquad
\dot{\mathbf e}=(\bm\Omega_n+\bm\omega_{\mathrm{rel}})\times\mathbf e.
\label{eq:odes}
\end{equation}
Equations~\eqref{eq:omega-n}--\eqref{eq:odes} are the overdamped,
mechanical counterpart of the models used for single-domain
nanoparticles with finite anisotropy, in which the moment turns inside
the particle against internal (N\'eel or Gilbert) damping while the
particle turns against the carrier liquid: the egg model of Shliomis and
Stepanov \cite{shliomis1993} and its extensions to rotating and
precessing fields \cite{usov2012,lyutyy2019}. At the nanoscale thermal
noise and the gyromagnetic precession of the moment enter, and the
internal motion cannot be observed. Here the internal degree of freedom
is a pair of macroscopic magnets, the inner friction is mechanical, and
the shell axis is tracked directly, so the phase diagram can be followed
as a function of $\kappa$ and compared with experiment.

\subsection{Dimensionless form}

The field, Eq.~\eqref{eq:field}, rotates at $\omega=2\pi f$, and the
observables compared with the model are the angles $\Theta$ and
$\alpha$ of $\mathbf n$, Eqs.~\eqref{eq:thetafold} and \eqref{eq:alpha}.
The two frictions enter only through their ratio
\begin{equation}
\kappa=\gamma_{\mathrm{in}}/\gamma_{\mathrm{out}},
\label{eq:kappa}
\end{equation}
which is the only dimensionless parameter beyond those of the
instantaneous theory. With the scales of Eqs.~\eqref{eq:Ba} and
\eqref{eq:wa}, Eqs.~\eqref{eq:omega-n} and \eqref{eq:omega-rel} become
\begin{align}
\frac{\bm\Omega_n}{\omega_a}
&=\hat{B}_0\,(\mathbf e\times\hat{\mathbf B}),
\label{eq:Om-dim}\\
\frac{\bm\omega_{\mathrm{rel}}}{\omega_a}
&=\frac{1}{\kappa}
\bigl[\hat{B}_0\,(\mathbf e\times\hat{\mathbf B})
 +(\mathbf e\cdot\mathbf n)\,(\mathbf e\times\mathbf n)\bigr],
\label{eq:om-dim}
\end{align}
where $\hat{\mathbf B}=\mathbf B/B_0$.

\subsection{Limits of the friction ratio}
\label{sec:limits}

Large $\kappa$ locks $\mathbf e$ to $\mathbf n$, and the particle
behaves as a rigid dipole. Small $\kappa$ makes
$\bm\omega_{\mathrm{rel}}$ fast: $\mathbf e$ relaxes at each instant
to a local minimum of Eq.~\eqref{eq:energy}, switches when that
minimum disappears (rotational hysteresis), and the dynamics of
$\mathbf n$ reduces to the instantaneous Stoner--Wohlfarth model of
C\={\i}murs and C\={e}bers
\cite{stonerWohlfarth,cimurs2013PRE,cimurs2013PRE87}.

In synchronous rotation $\mathbf e$ is at rest relative to the shell,
$\bm\omega_{\mathrm{rel}}=0$, so the state and, as the numerics confirm,
its stability range do not depend on $\kappa$. Synchronous rotation
exists for $\hat\omega<\hat\omega_{\mathrm I}(\hat B_0)$, with
\begin{equation}
\hat\omega_{\mathrm I}=
\begin{cases}
\hat{B}_0\sqrt{1-\hat{B}_0^2}, & \hat{B}_0\le 1/\sqrt{2},\\[2pt]
1/2, & \hat{B}_0\ge 1/\sqrt{2}.
\end{cases}
\label{eq:precession-bound}
\end{equation}
The value $1/2$ is the largest rate at which the anisotropy torque can
turn the shell (Sec.~\ref{sec:experiment}). Beyond this limit the
instantaneous theory has three regimes (Ref.~\cite{cimurs2013PRE} and
the $\kappa=0.02$ panel of Fig.~\ref{fig:phase_kappa}): conical
precession for $\hat B_0<1/2$, below the smallest switching field of
the pair; bistability of precession and planar libration for
$1/2<\hat B_0<1/\sqrt2$; and planar libration alone for
$\hat B_0>1/\sqrt2$. The precession cone narrows toward the rotation
axis as $\hat\omega$ grows, but it exists only below
$\hat B_0=1/\sqrt2$: for strong fields the instantaneous theory has no
attractor out of the field plane. Run d lies at $\hat B_0\approx1.1$
with the calibrated field scale of Sec.~\ref{sec:calibration} and at
$\hat B_0\approx1.5$ with the estimate of Eq.~\eqref{eq:Ba}, well above
this bound in either case. Its out-of-plane cone therefore requires a
finite $\kappa$.

\subsection{Symmetry and friction ratio}
\label{sec:symmetry}

With $B_z=0$, Eqs.~\eqref{eq:omega-n}--\eqref{eq:odes} are invariant
under $(n_z,e_z)\mapsto(-n_z,-e_z)$, and $n_z=e_z=0$ is an invariant
subspace. A scalar $\kappa$ therefore cannot select a pole in a
compensated circular field: cones about $+z$ and $-z$ are mirror
images, and which one is reached depends on the initial condition.
This is why the folded tilt, Eq.~\eqref{eq:thetafold}, is reported. The
invariant plane is left through the slightly tilted initial conditions
of Sec.~\ref{sec:numerics}; no bias field is applied.

A Stokes estimate of the inner friction, for two spheres of radius
$a=\SI{1.5}{\milli\metre}$ with the concentric-cavity factor
$(1-\lambda^3)^{-1}$ at clearance \SI{0.2}{\milli\metre}
($\lambda\simeq 0.88$) \cite{happel1983}, gives
$\gamma_{\mathrm{in}}\approx\SI{5e-7}{\joule\second}$ and
$\kappa\simeq 0.05$, near the instantaneous limit. The value
$\kappa=0.9$ used below is obtained by calibration
(Sec.~\ref{sec:calibration}); it is a model parameter, not that Stokes
prediction, and its physical meaning is discussed in
Sec.~\ref{sec:discussion}.

\section{Numerical protocol}
\label{sec:numerics}

We integrate Eq.~\eqref{eq:odes}, with the angular velocities in the
dimensionless form of Eqs.~\eqref{eq:Om-dim} and \eqref{eq:om-dim},
in the reduced time $\tau=\omega_a t$. The field direction is
\begin{equation}
\hat{\mathbf B}(\tau)=\bigl(\cos \hat{\omega}\tau,\,\sin \hat{\omega}\tau,\,0\bigr).
\label{eq:Bhat-num}
\end{equation}
This is the scaling of C\={\i}murs and C\={e}bers
\cite{cimurs2013PRE}. A run is therefore fixed by
$(\hat{\omega},\hat{B}_0,\kappa)$ and the initial state. To simulate a
measured point $(B_0,f)$ we take $\hat{\omega}=2\pi f/\omega_a$ and
$\hat{B}_0=B_0/B_a$ with the calibrated scales of
Sec.~\ref{sec:calibration}.

The six Cartesian components of $\mathbf n$ and $\mathbf e$ are
advanced with the adaptive Dormand--Prince method RK45
\cite{dormand1980}. The relative and absolute tolerances are
$2\times10^{-6}$ and $10^{-8}$. Both vectors are renormalized to unit
length at every evaluation step, and the step never
exceeds $8\%$ of the field period $2\pi/\hat{\omega}$. All averages
are taken over the last $35\%$ of each trajectory.
Each cell of a $22\times 14$ grid covering $0<\hat\omega\le2$ and
$0<\hat B_0\le1.2$ is started twice: from a near-equatorial orientation
$\mathbf n_0\propto(1,0,\tan 8^\circ)$ and from a near-polar orientation
$\mathbf n_0\propto(0.15,0.05,1)$, with $\mathbf e_0$ a small
perturbation of $\mathbf n_0$.

States are classified by the late-time mean tilt $\langle\Theta\rangle$,
Eq.~\eqref{eq:thetafold}, and the slip $s$ of Sec.~\ref{sec:observables}.
A state is planar when $|\langle\Theta\rangle-90^\circ|\le 10^\circ$.
Synchronous rotation (I) is planar with $s=0$, and libration (III) is
planar with $s>0$. Precession (II) is a cone,
$\langle\Theta\rangle<80^\circ$, locked to the field, $s=0$. A cell
is bistable (IV) when the two initial conditions end in II and III.
Cells in which a run does not settle to a steady state, a wobbling cone
or a tilted libration, are marked separately. The edge of region I is
drawn from Eq.~\eqref{eq:precession-bound}.

The norm drift stayed below $2\times10^{-4}$ in the stiff
small-$\kappa$ runs at measured points and below $2\times10^{-5}$ in
the range where cones form. The geometric thresholds are much wider
than these errors. The classification rests on the late-time behavior
of two trajectories per cell; it is not a Floquet stability analysis
and can miss narrow basins of attraction.

\begin{figure*}[!tp]
  \centering
  \includegraphics[width=\textwidth]{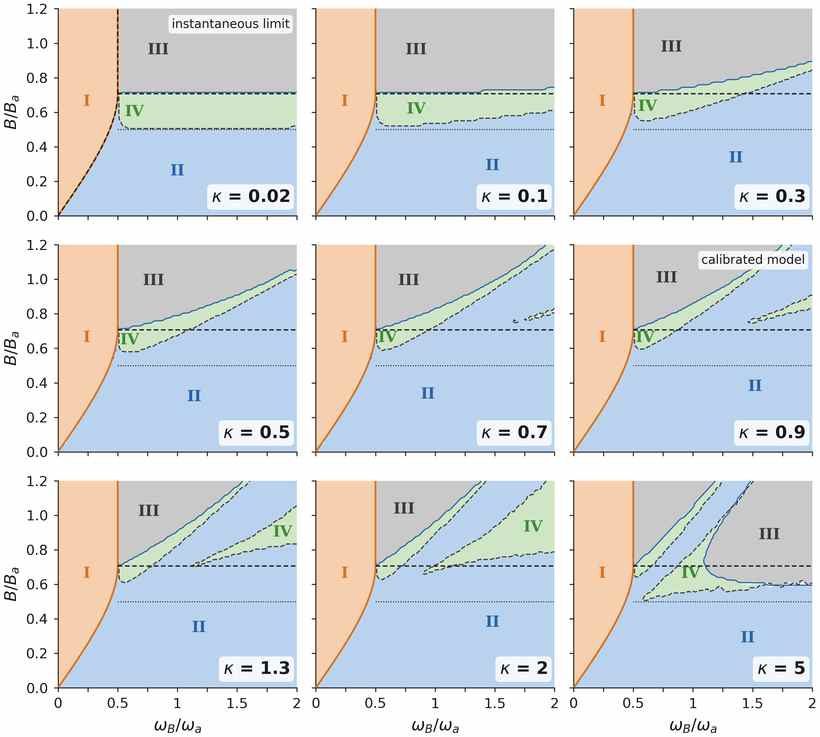}
  \caption{%
    Phase diagram of the two-friction model for nine values of the
    friction ratio $\kappa=\gamma_\mathrm{in}/\gamma_\mathrm{out}$, in the
    plane of reduced frequency $\hat\omega=\omega_B/\omega_a$ and reduced
    field $\hat B_0=B/B_a$; with the calibrated scales, $\hat\omega=1$ is
    \SI{4.5}{\hertz} and $\hat B_0=1$ is \SI{14.5}{\milli\tesla}. Colours
    mark the stable states found by linear stability on a $100\times80$
    grid: (I)~synchronous rotation in the field plane, (II)~precession on
    a cone locked to the field, (III)~in-plane libration with slips of
    the lag, and (IV)~bistability, where the cone and the libration are
    both stable and the initial condition selects between them. Dotted
    hatching ($\kappa=5$ only) marks cells without a stable fixed state.
    The solid orange curve is the edge of region~I,
    Eq.~\eqref{eq:precession-bound} for $\hat B_0<1/\sqrt2$ and
    $\hat\omega=1/2$ above; it is exact and the same for every $\kappa$.
    The black dashed line at $\hat B_0=1/\sqrt2$ is the boundary between
    regions III and IV of the instantaneous theory \cite{cimurs2013PRE},
    and the dotted line at $\hat B_0=1/2$ is the smallest field at which
    the moment switches. In the $\kappa=0.02$ panel all boundaries of the
    instantaneous theory are dashed; at this $\kappa$ the model is close
    to the instantaneous limit. The $\kappa=0.9$ panel is the calibrated
    model of Fig.~\ref{fig:phasemap}.
  \label{fig:phase_kappa}}
\end{figure*}

\section{Results}
\label{sec:results}

\subsection{Inner friction and the phase diagram}
\label{sec:kappa}

Figure~\ref{fig:phase_kappa} shows how the phase diagram of the
two-friction model changes with the friction ratio. At $\kappa=0.02$ it
reproduces the instantaneous theory: region I is bounded by
Eq.~\eqref{eq:precession-bound}, the bistable band IV fills
$1/2<\hat B_0<1/\sqrt2$ for $\hat\omega>1/2$, and only libration remains
above $\hat B_0=1/\sqrt2$. The edge of region I is the same in every
panel. At $\kappa=0.1$ the map is still close to the instantaneous one,
apart from a slight rise of the lower edge of the bistable band.

From $\kappa\approx0.3$ the bistable band tilts. Its upper edge, above which only
libration remains, rises with frequency, to $\hat B_0\approx0.9$ at
$\hat\omega=2$ for $\kappa=0.3$ and to $\hat B_0\approx1.05$ for
$\kappa=0.5$, while the bistable band narrows. Below the tilted band, region II
extends above $\hat B_0=1/\sqrt2$: the particle precesses at fields
where the instantaneous theory permits only libration. For
$\kappa\ge0.7$ the upper edge leaves the plotted range,
$\hat B_0=1.2$, at $\hat\omega\approx1.95$ ($\kappa=0.7$), $1.75$
($\kappa=0.9$), $1.6$ ($\kappa=1.3$), and $1.4$ ($\kappa=2$). At the
same time a second bistable tongue appears at high frequency inside the
extended region II; it widens and moves to lower frequency as $\kappa$
grows. At $\kappa=5$ libration returns: a second region III occupies
high frequencies above $\hat B_0\approx0.6$, and precession above
$\hat B_0=1/\sqrt2$ survives only in a narrow channel. Precession at
strong fields therefore requires an inner friction that is neither
negligible nor dominant.

The cones in the extended region II lie close to the rotation axis. At
$\kappa=0.9$ and $\hat\omega=1.9$ the model settles on cones with
$\Theta$ between about \ang{12} at $\hat B_0=0.4$ and \ang{23} at
$\hat B_0=1.0$. In the instantaneous limit, cones of similar latitude
($\Theta\approx$~\ang{9}--\ang{12} at $\hat\omega=1.9$) exist only for
$\hat B_0<1/\sqrt2$. The polar phase at strong fields is thus the
high-frequency precession of the instantaneous theory, carried across
the bound $\hat B_0=1/\sqrt2$ by the finite inner friction.

\subsection{Calibration against the experiment}
\label{sec:calibration}

\begin{figure}[!htbp]
  \centering
  \includegraphics[width=0.92\columnwidth]{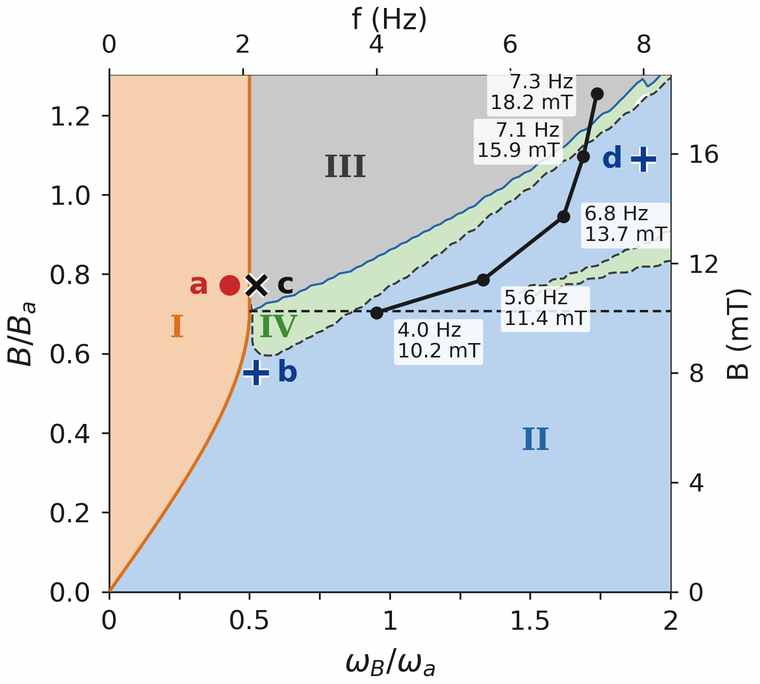}
  \caption{Calibrated two-friction model ($\kappa=0.9$,
$B_a=\SI{14.5}{\milli\tesla}$, $\omega_a=\SI{26.4}{\per\second}$)
compared with experiment. Shading shows the stable states of the
model, regions I--IV coloured as in Fig.~\ref{fig:phase_kappa}. The
orange line is the edge of region~I, Eq.~\eqref{eq:precession-bound},
and the dashed line the boundary between III and IV of the
instantaneous theory, $\hat B_0=1/\sqrt2$. Symbols mark the four runs
of Table~\ref{tab:runs} and Fig.~\ref{fig:regimes} by their observed
state: red circle, synchronous rotation (a, \SI{11.2}{\milli\tesla},
\SI{1.8}{\hertz}); blue plus, precession (b, \SI{8.0}{\milli\tesla},
\SI{2.2}{\hertz}, and d, polar precession, \SI{15.8}{\milli\tesla},
\SI{8}{\hertz}); black cross, libration (c, \SI{11.2}{\milli\tesla},
\SI{2.2}{\hertz}). The black line with points is the measured edge of
region~III, each point labelled with its drive frequency and field.
The top and right axes give laboratory units.}
\label{fig:phasemap}
\end{figure}

\begin{figure*}[!tp]
\centering
\includegraphics[width=0.82\textwidth]{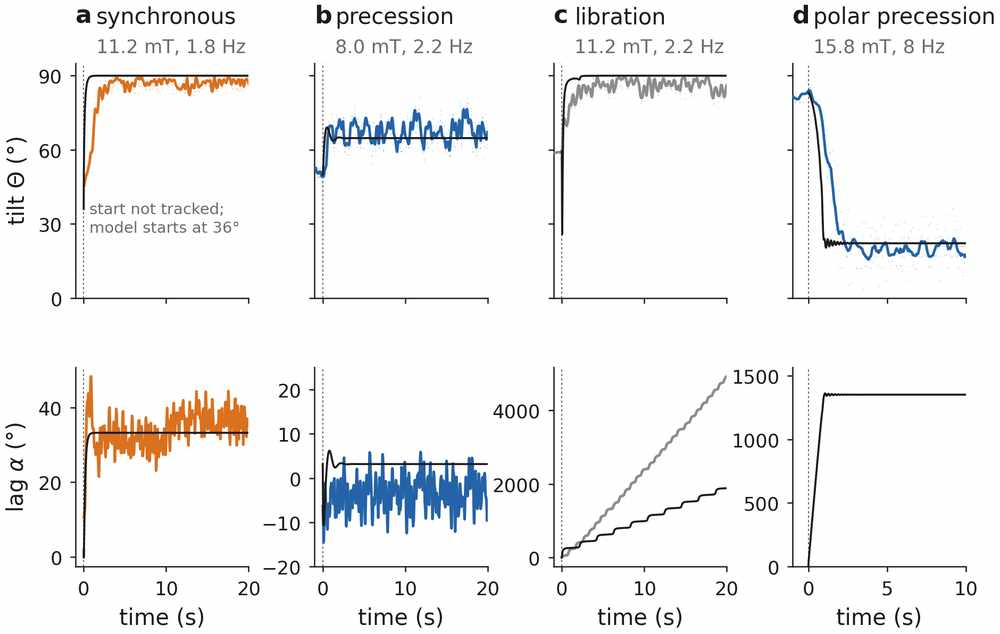}
\caption{Four measured records (colored by regime) and the two-friction
model (black; $B_a=\SI{14.5}{\milli\tesla}$,
$\omega_a=\SI{27.0}{\per\second}$, $\kappa=0.9$), against time after
field onset (dotted line). Top row: tilt $\Theta=\Theta_{\mathrm{fold}}$,
Eq.~\eqref{eq:thetafold}. Bottom row: accumulated lag
$\alpha$, Eq.~\eqref{eq:alpha}, set to zero at the first tracked frame
after onset. Thick color lines are experimentally measured angles
\textbf{a}, Synchronous, \SI{11.2}{\milli\tesla}, \SI{1.8}{\hertz}: the
start tilt and the first \SI{0.3}{\second} are not tracked, and the
model starts at an estimated $36^\circ$ (plausible range
\SIrange{32}{45}{\degree}); $\Theta$ measured $87^\circ$, model
$90^\circ$; slip 0 in both.
\textbf{b}, Precession (locked tilted cone), \SI{8.0}{\milli\tesla},
\SI{2.2}{\hertz}: $\Theta$ measured $67^\circ$, model $65^\circ$; slip 0
in both.
\textbf{c}, Libration, \SI{11.2}{\milli\tesla}, \SI{2.2}{\hertz}:
$\Theta$ measured $86^\circ$, model $90^\circ$; slip $s$ (late-time slope
of $\alpha$ divided by $360^\circ f$) measured $0.30$, model $0.11$.
\textbf{d}, Polar precession, \SI{15.8}{\milli\tesla}, \SI{8}{\hertz}: $\Theta$ measured
\SIrange{20}{24}{\degree} (rim-cap corrected; the plotted trace is the
uncorrected estimator, which settles at about $19.5^\circ$), model
$22^\circ$, locked. The measured lag cannot be
followed through the rise to the pole.}
\label{fig:regimes}
\end{figure*}

\begin{table*}[!tp]
\caption{Measured runs and the calibrated two-friction model
($\kappa=0.9$, $B_a=\SI{14.5}{\milli\tesla}$). Reduced coordinates use
$\omega_a=\SI{26.4}{\per\second}$, as in Fig.~\ref{fig:phasemap}; model
tilts and slips are taken from the time series of
Fig.~\ref{fig:regimes}, computed with $\omega_a=\SI{27.0}{\per\second}$.
Fields are coil settings converted at \SI{1.6}{\milli\tesla} per unit.
The tilt of run d is rim-cap corrected; its slip could not be measured
at 25 frames per second.}
\label{tab:runs}
\begin{ruledtabular}
\begin{tabular}{llcccccccc}
Run & Observed state & $B_0$ (mT) & $f$ (Hz) & $\hat{\omega}$ & $\hat{B}_0$
 & $\Theta$ meas. & $\Theta$ model & $s$ meas. & $s$ model\\
\hline
a & Synchronous & 11.2 & 1.8 & 0.43 & 0.77 & \ang{87} & \ang{90} & 0 & 0\\
b & Precession & 8.0 & 2.2 & 0.52 & 0.55 & \ang{67} & \ang{65} & 0 & 0\\
c & Libration & 11.2 & 2.2 & 0.52 & 0.77 & \ang{86} & \ang{90} & 0.30 & 0.11\\
d & Polar precession & 15.8 & 8.0 & 1.90 & 1.09 & \ang{20}--\ang{24} & \ang{22} & -- & 0\\
\end{tabular}
\end{ruledtabular}
\end{table*}

Runs a and c have the same field, which lies above
$1/\sqrt2$ for any $B_a<\SI{15.8}{\milli\tesla}$, and bracket the
vertical edge $\hat\omega=1/2$ of region I. Because that edge does not
depend on $\kappa$, the two runs confine $\omega_a$ to
$4\pi\times\SI{1.8}{\hertz}<\omega_a<4\pi\times\SI{2.2}{\hertz}$, that
is, to \SIrange{22.6}{27.6}{\per\second}. Figure~\ref{fig:phasemap}
uses $\omega_a=\SI{26.4}{\per\second}$, and the time series of
Fig.~\ref{fig:regimes} use $\omega_a=\SI{27.0}{\per\second}$, both
inside this window.

The remaining two scales follow from run d. At
$\omega_a=\SI{27.0}{\per\second}$ and $B_a=\SI{14.5}{\milli\tesla}$ the
model reaches a polar cone from a start near the field plane, as the
particle does, for $0.7\lesssim\kappa\lesssim1.6$; at $\kappa=0.6$ and at
$\kappa=2$ it librates instead. Across this window the cone angle grows
from \ang{18.5} to \ang{30}, and the measured \SIrange{20}{24}{\degree}
selects $0.8\lesssim\kappa\lesssim1.0$. At $\kappa=0.9$ the cone angle
changes only from \ang{20} to \ang{23} as $B_a$ varies from
\SI{13.5}{\milli\tesla} to \SI{16}{\milli\tesla}; for smaller $B_a$ a
start near the plane ends in libration, and for $B_a>\SI{15.8}{\milli\tesla}$
runs a and c would fall below $\hat B_0=1/\sqrt2$. We use $\kappa=0.9$
and $B_a=\SI{14.5}{\milli\tesla}$ inside these ranges. The tilt of run d
is therefore fitted, not predicted. What the calibration does not
supply is the existence of the cone: the instantaneous limit has no
out-of-plane attractor at this field for any admissible scale, whereas
the two-friction model has one for $\kappa$ of order unity, and the same
parameter set reproduces the other three runs.

Figure~\ref{fig:phasemap} places the four runs on the calibrated map,
and Table~\ref{tab:runs} lists their coordinates. Runs a and c sit on
either side of the vertical edge of region I, at $\hat B_0=0.77$; run b
lies in region II just below the bistable band; and run d, at
$\hat\omega=1.90$ and $\hat B_0=1.09$, lies in the part of region II
that extends above $\hat B_0=1/\sqrt2$. The model reproduces the
observed state of all four runs. In the instantaneous limit,
$\kappa=0.02$, the same working point d librates in the field plane
with slip $s=0.93$.

The black line in Fig.~\ref{fig:phasemap} is the measured edge of
region III, recorded between \SI{4.0}{\hertz} and \SI{7.3}{\hertz}. Its
field rises from \SI{10.2}{\milli\tesla} to \SI{18.2}{\milli\tesla}, by
a factor of $1.8$ that does not depend on the coil constant. The
instantaneous theory predicts a frequency-independent edge at
$\hat B_0=1/\sqrt2$; the model at $\kappa=0.9$ predicts an edge that
rises with frequency, as observed, although less steeply. From
\SI{4.0}{\hertz} to \SI{6.8}{\hertz} the measured edge lies below the
model edge, in region II, and between \SI{7.1}{\hertz} and
\SI{7.3}{\hertz} it turns up and crosses it. The field of the edge is
converted from coil units at \SI{2.28}{\milli\tesla} per unit, a factor
fixed by placing its lowest point on the instantaneous edge and different
from the nominal \SI{1.6}{\milli\tesla} per unit used for runs a--d. With
the nominal factor the edge would lie \SI{30}{\percent} lower in
$\hat B_0$. The comparison therefore concerns the shape of the edge,
whose rise by a factor of $1.8$ does not depend on this factor, rather
than its absolute position.

\subsection{Time series of the four regimes}
\label{sec:timeseries}

Figure~\ref{fig:regimes} compares the four records with the model, in
tilt $\Theta$ (top row) and accumulated lag $\alpha$ (bottom row); the
values are collected in Table~\ref{tab:runs}. At \SI{11.2}{\milli\tesla}
and \SI{1.8}{\hertz} [Fig.~\ref{fig:regimes}(a)] both are synchronous:
$\Theta$ settles in the field plane and $\alpha$ levels off at a
constant lag. The model reaches the plane faster than the record. At
\SI{8.0}{\milli\tesla} and \SI{2.2}{\hertz} [Fig.~\ref{fig:regimes}(b)]
both form a locked tilted cone, at $\Theta=\ang{67}$ measured and
\ang{65} in the model, with a flat lag. At \SI{11.2}{\milli\tesla} and
\SI{2.2}{\hertz} [Fig.~\ref{fig:regimes}(c)] both librate in the plane,
and $\alpha$ grows in steps, one per slipped turn. The model reproduces
the regime but slips about three times more slowly, $s=0.11$ against
$0.30$ measured. At \SI{15.8}{\milli\tesla} and \SI{8}{\hertz}
[Fig.~\ref{fig:regimes}(d)] both leave the plane and lock onto a cone
near the pole, at \SIrange{20}{24}{\degree} measured and \ang{22} in
the model. Before it locks, the model slips about four turns in the
plane, and the measured particle leaves the plane later than the model.

\subsection{Structure of the polar state}
\label{sec:polar}

The model shows how the polar state is maintained at run d. The shell
axis stands \ang{22} from the rotation axis, and its projection on the
field plane is perpendicular to the field. The magnetic direction
$\mathbf e$ turns with the field at an azimuthal lag of about \ang{55}
and a tilt of about \ang{23} out of the field plane, and it keeps a
fixed angle of about \ang{50} to the easy axis $\pm\mathbf n$. At this
angle the field and anisotropy torques on the inner pair do not
balance. Their resultant, Eq.~\eqref{eq:om-dim}, drives a steady
rotation of the magnets relative to the shell,
$|\bm\omega_{\mathrm{rel}}|\approx0.9\,\omega_a$, against the inner
friction. In the instantaneous limit, $\kappa\to0$, such a torque
imbalance relaxes at once and $\mathbf e$ returns to an energy minimum,
so the state cannot exist. In the opposite limit, $\kappa\to\infty$,
$\mathbf e$ is locked to $\mathbf n$ and the relative rotation is
suppressed, consistent with the return of libration at run d for
$\kappa=2$ (Sec.~\ref{sec:calibration}) and across high frequencies at
$\kappa=5$ in Fig.~\ref{fig:phase_kappa}.

\section{Discussion and conclusion}
\label{sec:discussion}

A single egg-yolk particle shows every orbit class of a magnetic body in
a rotating field: synchronous rotation, precession, and libration. At
high drive it also shows a polar phase that the instantaneous theory
rules out. That theory holds the body in the field plane whenever
$\hat B_0>1/\sqrt2$. At \SI{15.8}{\milli\tesla} and \SI{8}{\hertz}, well
above this bound with either the calibrated or the a priori field scale,
the shell instead leaves the plane and locks onto a cone only
\SIrange{20}{24}{\degree} from the rotation axis. To our knowledge, this
is the first observation of a high-drive polar state for a particle
whose magnetic moment can move inside it.

The two-friction model explains this state by adding one physical
ingredient: the inner magnets follow the field with a finite delay.
With one parameter set, $B_a=\SI{14.5}{\milli\tesla}$, $\kappa=0.9$,
and $\omega_a=\SI{27.0}{\per\second}$, and without a bias field, it reproduces the terminal state of all four measured
records (Table~\ref{tab:runs}) and matches their tilt within \ang{4}:
\ang{87} measured against \ang{90} modeled for synchronous rotation,
\ang{67} against \ang{65} for precession, \ang{86} against \ang{90} for
libration, and \SIrange{20}{24}{\degree} against \ang{22} for the polar
cone. These differences are comparable to the measurement uncertainty;
for the polar record alone, the tilt estimate spans \ang{4} after the
rim-cap correction. The polar state is the central result. The
instantaneous theory has no such state at the measured field and
frequency, whereas the two-friction model has one for any $\kappa$
between about $0.7$ and $1.6$; the measured tilt then fixes $\kappa$.
In the model the state is carried by a steady rotation of the magnets
inside the shell (Sec.~\ref{sec:polar}), which is precisely the degree of
freedom that the instantaneous theory leaves out.

The model does not match every detail, and the differences are
informative. In libration the model slips about three times more slowly
than the particle, $s=0.11$ against $0.30$. Matching the measured slip
would require $\omega_a\approx\SI{22}{\per\second}$, below
\SI{22.6}{\per\second}, where run a would no longer rotate
synchronously. Close to the synchronous edge the slip is steep in
$\omega_a$: a reduction of $\omega_a$ by \SI{15}{\percent} raises it
from $0.11$ to $0.28$. The mismatch therefore reflects the steepness of
the dynamics near that boundary, not a failure to predict the regime. The
transients differ as well. The model reaches the field plane faster
than the record in Fig.~\ref{fig:regimes}(a), and the measured polar
trajectory leaves the plane later than the modeled one. The overdamped
model neglects fluid and shell inertia, which reach
$\mathrm{Re}\sim1$ and $I_s\omega/\gamma_{\mathrm{out}}\approx0.15$ at \SI{8}{\hertz}
(Sec.~\ref{sec:observables}), and the hydrodynamic interaction with
the nearby cavity walls; all three would delay the response.

The calibrated scales can be compared with the a priori estimates of
Sec.~\ref{sec:experiment}. The anisotropy energy $KV=mB_a=\SI{3.5e-4}{\joule}$
is $1.4$ times the dipolar estimate, and the drag coefficient
$\gamma_{\mathrm{out}}=mB_a/\omega_a=\SI{1.3e-5}{\joule\second}$ is $1.3$ times the
hydrodynamic estimate. Accordingly, $\omega_a$ agrees with its estimate
within \SI{10}{\percent}, and $B_a$ exceeds its estimate by
\SI{40}{\percent}. Both deviations lie within the systematic
uncertainties discussed in Sec.~\ref{sec:observables}. The inner
friction is different. The calibrated
$\gamma_{\mathrm{in}}=\kappa\gamma_{\mathrm{out}}\approx\SI{1.2e-5}{\joule\second}$
is about 20 times the Stokes estimate for spheres turning freely in
their cavities. A plausible reason is that the two magnets attract each
other, with a force $3\mu_0m_0^2/(2\pi d^4)\approx\SI{0.04}{\newton}$,
some 30 times the weight of a magnet, and press against the printed
walls, so that their rotation is resisted by contact rather than by the
glycerol film alone. Because all three scales are calibrated rather than
measured independently, the comparison shows that the model is
consistent with the data. It is not yet an independent prediction of
the phase boundaries.

The next step is to turn the calibration into a prediction. The follow-up experiment would measure the three-axis
field at the particle, record the shell from two synchronized views,
and track the internal magnets directly. Those data would fix $B_a$,
$\gamma_{\mathrm{in}}$, and the field scale of the edge series without
calibration. They would also separate a
steady cone from a large-amplitude oscillation and map where the polar
phase begins across the $(\hat\omega,\hat B_0)$ plane.



\bibliographystyle{apsrev4-2}
\bibliography{references}

\end{document}